\documentclass[aps,prl,twocolumn,longbibliography,showpacs,floatfix]{revtex4-1}

\usepackage{graphicx}
\usepackage{amsmath,amssymb,bbold,bm,color}
\usepackage{float}
\usepackage{epstopdf}
\usepackage{hyperref}
\usepackage{color}
\usepackage{enumerate}
\usepackage{wasysym}
\usepackage{url}
\usepackage{dsfont}
\usepackage{braket}

\newcommand{\up}{\uparrow}
\newcommand{\down}{\downarrow}

\hypersetup{colorlinks=true,linkcolor=blue,citecolor=blue,urlcolor=blue}

\begin{document}

\title{Superconducting instabilities in a spinful Sachdev-Ye-Kitaev model}

\author{\'Etienne Lantagne-Hurtubise}
\affiliation{Department of Physics and Astronomy \& Stewart Blusson Quantum Matter Institute, University of British Columbia, Vancouver BC, Canada V6T 1Z4}

\author{Vedangi Pathak}
\affiliation{Department of Physics and Astronomy \& Stewart Blusson Quantum Matter Institute, University of British Columbia, Vancouver BC, Canada V6T 1Z4}

\author{Sharmistha Sahoo}
\affiliation{Department of Physics and Astronomy \& Stewart Blusson Quantum Matter Institute, University of British Columbia, Vancouver BC, Canada V6T 1Z4}

\author{Marcel Franz}
\affiliation{Department of Physics and Astronomy \& Stewart Blusson Quantum Matter Institute, University of British Columbia, Vancouver BC, Canada V6T 1Z4}

\date{\today}

\begin{abstract}
We introduce a spinful variant of the Sachdev-Ye-Kitaev model with an effective time reversal symmetry, which can be solved exactly in the limit of a large number $N$ of degrees of freedom.
At low temperature, its phase diagram includes a compressible non-Fermi liquid and a strongly-correlated spin singlet superconductor that shows a tunable enhancement of the gap ratio predicted by BCS theory. These two phases are separated by a first-order transition, in the vicinity of which a gapless superconducting phase, characterized by a non-zero magnetization, is stabilized upon applying a Zeeman field. We study equilibrium transport properties of such superconductors using a lattice construction,
and propose a physical platform based on topological insulator flakes where they may arise from repulsive electronic interactions.
\end{abstract}
\maketitle

Understanding strongly correlated forms of superconductivity, going beyond the celebrated BCS~\cite{Bardeen1957a, Bardeen1957b, Gorkov1958} and Migdal-Eliashberg~\cite{Migdal1958, Eliashberg1960, Nambu1960, Marsiglio2020} theories, remains an ongoing avenue of research. One of the main difficulties lies in the rarity of tractable models~\cite{Phillips2020, Liu2021, Crepel2020} providing analytical insight into this phenomenon.
Recently, the advent of exactly-solvable models of non-Fermi liquids, the family of so-called SYK models~\cite{SY1993, Sachdev2015, Kitaev2015, Maldacena2016}, has sparked remarkable progress in exploring correlated phases with intriguing properties such as strange metallic transport and maximal chaos~\cite{Banerjee2017, Jian2017, Gu2017, Song2017, Bi2017, Patel2018a, Chowdhury2018, Wu2018, Cha2020, Kim2020}. Solvable models of correlated superconductors have been similarly constructed -- two popular approaches consisting of explicitly adding pairing terms to an SYK construction~\cite{Patel2018b, Cheipesh2019, Wang2020b} or considering random Yukawa electron-phonon interactions~\cite{Esterlis2019, Wang2020a, Wang2020c, Hauck2020, Pan2021}. 

Building on these ideas, in this work we introduce a simple model for correlated superconductivity with rich phenomenology, where the superconducting correlations are instead generated directly by disordered SYK-type fermionic interactions~\cite{Chowdhury2020a, Chowdhury2020b}. It consists of a pair of coupled complex SYK (cSYK) models~\cite{SY1993, Sachdev2015, Gu2020} with random two-body interactions that are constrained by an anti-unitary time reversal symmetry, and can thus be regarded as a \emph{spinful} generalization of the SYK model. This is inspired by recent work on a related but subtly different symmetry setting, where two SYK models are instead related by a unitary symmetry~\cite{maldacena2018,Klebanov2019, Sahoo2020, Klebanov2020, Garcia-Garcia2021}, and which hosts both (gapped) symmetry-broken and (gapless) non-Fermi liquid phases with a holographic interpretation. 

\begin{figure}
\centering
\includegraphics[width=0.38\columnwidth]{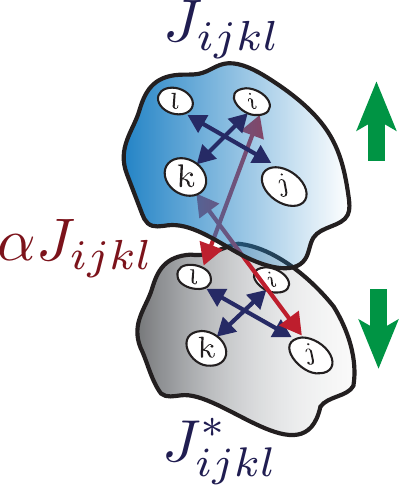}
\includegraphics[width=0.6\columnwidth]{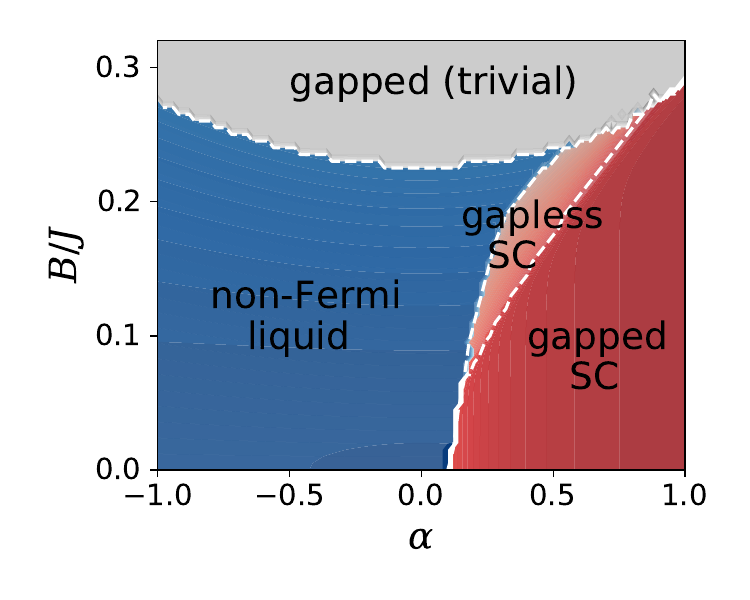}
    \caption{(Left:) Illustration of the coupling terms in the spinful SYK model, Eq.~\ref{eq:H_two_TR_cSYKS}. 
    (Right:) Low temperature ($\beta J = 100$) phase diagram as a function of Zeeman field $B$ and interaction parameter $\alpha$, at charge neutrality $\mu=0$. For $\alpha < 0$ the SYK non-Fermi liquid is stable, whereas for $\alpha > 0$ we find an instability to a gapped spin-singlet superconductor. Interestingly a region of gapless superconductivity with finite magnetization is stabilized at non-zero $B$. White dashed lines denote first-order phase transitions.
    }
    \label{fig1}
\end{figure}

In analogy with the results of these works, at low temperature the spinful SYK model shows the spontaneous breaking of a U(1) symmetry. However, rather than the breaking of an \emph{axial} U(1) symmetry leading to a ``traversable wormhole" phase~\cite{Sahoo2020, Klebanov2020, Garcia-Garcia2021}, the \emph{global} U(1) symmetry is instead broken, driving the system to a correlated spin-singlet superconducting phase. This superconductor shows an enhanced gap ratio compared to the BCS prediction, and might also exhibit connections to holography. It is separated by a first-order transition from an SYK non-Fermi liquid, in the vicinity of which a \emph{gapless} superconducting phase, characterized by a finite magnetization, is stabilized upon applying a Zeeman field $B$ (see a schematic low-temperature phase diagram in Fig.~\ref{fig1}). Using a lattice construction with spinful SYK models at each site we compute the equilibrium transport properties of the two SC phases, finding sharp qualitative differences in their supercurrent-phase relations.

\emph{The model. --} We consider a variant of the SYK model that consists of a (0+1)-dimensional  ``quantum dot" with a large number $N$ of degrees of freedom, each coming in two flavors $a= \up, \down$. We assume all-to-all, random interactions between degrees of freedom of the same flavor, described by the complex SYK Hamiltonian
\begin{equation}
    H_a = \sum_{ijkl=1}^N J^a_{ij;kl} c_{i a}^\dagger c_{j a}^\dagger c_{k a} c_{l a} - \mu_a \sum_j c^\dagger_{j a} c_{j a} ,
    \label{eq:H_cSYK}
\end{equation}
where the coupling constants are drawn from a Gaussian distribution with zero mean and variance $ \overline{ |J^a_{ijkl}|^2} = \frac{J^2}{8 N^3}$, and $\mu_a$ are chemical potentials that can be tuned independently for the two species. Fermionic commutation relations impose the constraints $J^a_{ij;kl} = - J^a_{ij;lk} = -J^a_{ji;kl} = (J^a_{kl;ij})^*$ on the coupling constants. In the following we also impose the stronger requirement that $J^a_{ij;kl}$ be fully anti-symmetric~\footnote{\label{footnote1}In the large $N$ limit such fully anti-symmetric terms are expected to dominate as their number scales as $\sim N^4$, rather than $N^3$ or $N^2$ for interaction terms with one or two repeated indices. In mesoscopic systems, these omitted terms could however play an important role.}. We then require invariance under the anti-unitary symmetry $\Theta= \tau^x \mathcal{K}$, where $\tau^x$ is a Pauli matrix acting on the flavor degree of freedom and $\mathcal{K}$ denotes complex conjugation. This
enforces $J^\up_{ij;kl} = ( J_{ij;kl}^\down)^* = J^{\down}_{kl;ij}$.

We now couple the cSYK models with two-body interactions that conserve charge for each flavor (with U(1) $\otimes$ U(1) symmetry), of the form $ J^{a b}_{ijkl} c_{i a}^\dagger c_{j b}^\dagger c_{k a} c_{l b}$. Consistency with the anti-unitary symmetry requires that $ J_{ijkl}^{a b} = ( J_{ijkl}^{b a})^*$.
For concreteness we consider the coupling constants generated by Coulomb interactions in a degenerate manifold that is constrained
by $\Theta$ (see Appendix for details and connections to a proposed experimental platform based on a topological insulator flake). This enforces the constraints $J^{a b}_{il;kj} = \alpha J^a_{ij;kl} = \alpha J^b_{kl;ij} $,
with $\alpha$ a dimensionless constant controlling the ratio of inter to intra-flavor interactions. In the proposed physical platform $\alpha > 0$ ($\alpha < 0$) corresponds to repulsive (attractive) inter-flavor interactions. We thus consider 
\begin{align}
H =& \sum_{ijkl} J_{ij;kl} \Big[ c_{i \up}^\dagger c_{j \up}^\dagger c_{k \up} c_{l \up} +  c_{k \down}^\dagger c_{l \down}^\dagger c_{i \down} c_{j \down} \nonumber \\
& + \alpha \left( c_{i \up}^\dagger c_{l \down}^\dagger c_{k \up} c_{j \down} + c_{k \down}^\dagger c_{j \up}^\dagger c_{i \down} c_{l \up} \right)  \Big] \nonumber \\
- &(\mu + B) \sum_{j} c^\dagger_{j \up} c_{j \up} - ( \mu - B) \sum_{j } c^\dagger_{j \down} c_{j \down} 
\label{eq:H_two_TR_cSYKS},
\end{align}
where we expressed $\mu_{\up, \down} = \mu \pm B$ in terms of a (global) chemical potential $\mu$ and a Zeeman term $B$ which breaks the anti-unitary symmetry $\Theta$. For $\mu=0$ the Hamiltonian is invariant under the combination of flavor and particle-hole transformation $ c_{i a}^\dagger \leftrightarrow  c_{ib}$ with $a \neq b$.

\emph{Saddle-point equations.--} We first consider the charge neutrality point, $\mu=0$. The Euclidean-time path integral formulation of the model at inverse temperature $\beta = 1/k_B T$ reads $\mathcal{Z} = \int [\mathcal{D}[c, c^\dagger] e^{- S}$ with the effective action
$S = \int_0^\beta d\tau \left( \sum_{i,a} c^\dagger_{i a}(\tau) \partial_{\tau} c_{i a}(\tau) + H \right)$.
Averaging over quenched disorder in the couplings $J_{ijkl}$, and considering only replica-diagonal solutions (assuming no spin glass physics~\footnote{In the context of SYK physics, replica off-diagonal solutions are believed to be sub-leading in the large-N limit~\cite{Suh2018, Wang2019, Arefeva2019, Baldwin2020}. An extension of this analysis to the spinful SYK model is beyond the scope of the present work.}), we obtain an effective action written in terms of the (standard and anomalous) averaged Green's functions $G_{\tau, \tau'} = \frac{1}{N} \sum_j \langle \mathcal{T} c_{j \up}(\tau) c_{j \up}^\dagger(\tau') \rangle$ and $F_{\tau, \tau'} = \frac{1}{N} \sum_j \langle \mathcal{T} c_{j\up}(\tau) c_{j\down}(\tau') \rangle$ and their respective self-energies $\Sigma$ and $\Pi$ (see Appendix for details).
From this effective action the semiclassical ($N \rightarrow \infty$) saddle-point equations are obtained by taking functional derivatives with respect to the Green's functions and self-energies,
\begin{align}
\Sigma_\tau &= -J^2 \Big[ (1+\frac{\alpha^2}{2}) G^2_\tau G_{-\tau}  - 2 \alpha G_{\tau} F_{\tau} F_{-\tau} + \frac{\alpha^2}{2}  F^2_{\tau} G_{-\tau} \Big] \nonumber \\
\Pi_{\tau} &= -J^2 \Big[ (1+\frac{\alpha^2}{2}) F^2_{\tau} F_{-\tau} - 2 \alpha  F_{\tau} G_{\tau} G_{-\tau} + \frac{\alpha^2}{2} G^2_{\tau}  F_{-\tau} \Big] \nonumber \\
G_n &= -\frac{B + \Sigma_n + i \omega_n }{D_n} \quad , \quad
F_n = \frac{ \Pi_n}{D_n} ,
\label{eq:saddle_point_equations}
\end{align}
where $ D_n = (B + \Sigma_n + i \omega_n )^2 - \Pi^2_n$. Here we used time translation invariance to express $ G_{\tau,\tau'} \equiv G_{\tau-\tau'} $,
while $G_n \equiv G(\omega_n)$ (and similarly) are Fourier transformed expressions in terms of fermionic Matsubara frequencies $\omega_n = (2n + 1) \pi T$. 
\begin{figure*}
\includegraphics[width=\textwidth]{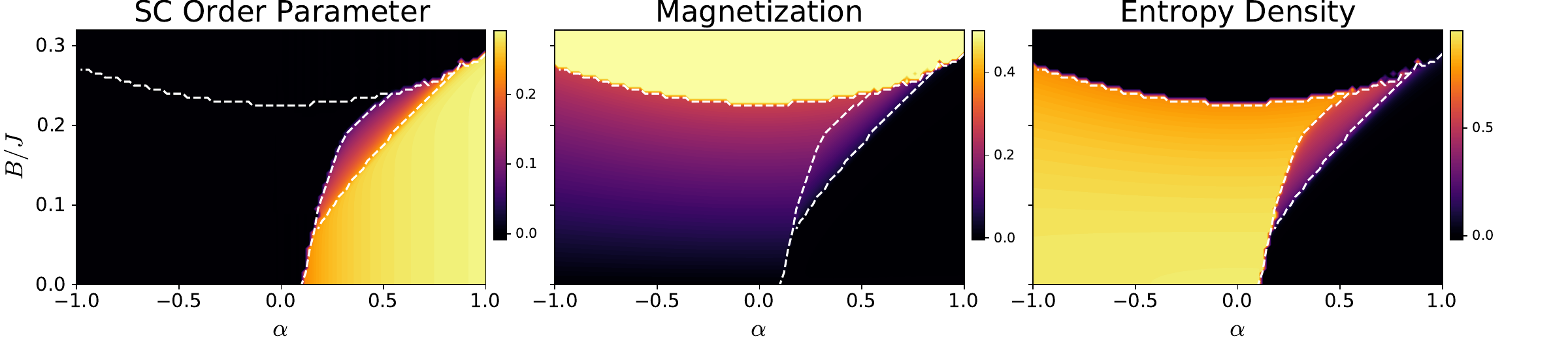}
    \caption{Phase diagram of the model [Eq.~(\ref{eq:H_two_TR_cSYKS})] at low-temperature $\beta J = 100$ and charge neutrality $\mu=0$. The superconducting order parameter $\Delta$ (left panel),  magnetization $M$ (middle panel) and residual entropy density $\mathcal{S}_0$ (right panel) are obtained from the self-consistent solutions of Eqs.~(\ref{eq:saddle_point_equations}) as a function of $\alpha$ and $B$. Dashed white lines indicate first-order phase transitions.
    \label{fig2}}
\end{figure*}
This set of coupled equations can be solved self-consistently through an iterative method until convergence is attained. In practice, as coupled models of this type~\cite{maldacena2018, Klebanov2019, Sahoo2020, Klebanov2020, Garcia-Garcia2021} often exhibit first-order phase transitions, we sweep the Zeeman field $B$ back and forth and feed the converged solution for the next value of $B$ considered. This gives rise to hysteresis curves from which one picks the solution with the lowest free energy density $ \mathcal{F} = -T \ln \mathcal{Z}/N$, given in the large $N$ limit by substituting the saddle point solutions in the action~\footnote{Here we regularized the free energy (divided by $T$) using its value for $2N$ non-interacting complex fermions, $2 \sum_{n} \ln(i\omega_{n}) = 2 \ln 2$, to regularize divergences at large frequencies in the numerical evaluation of $D_n$},
\begin{align} 
\frac{-\mathcal{F}}{T} &= 2 \ln 2 + \sum_{\omega_{n}} \left[ \ln \left( \frac{D_n}{(i\omega_{n})^2} \right) +\frac{3}{2}\left( \Sigma_n G_n +  \Pi_n F_n \right) \right] .
\label{eq:free_energy}
\end{align} 
Similarly, the entropy density $\mathcal{S} = \left( \mathcal{U} - \mathcal{F} \right)/T$ is obtained, with the energy density
\begin{align}
    \mathcal{U}
    = T \sum_{\omega_n} \left[ 2 B G_n +   \Sigma_n G_n + \Pi_n F_n \right],
\end{align} 
and  the magnetization $M = \frac{1}{2N} \sum_j \langle c^\dagger_{j \up} c_{j \up} - c^\dagger_{j \down} c_{j \down} \rangle$  can be read off from $M =  \frac{1}{2} - G_{\tau=0^+}$.

\emph{Phase diagram.--} We first explore the low-temperature physics of the model by self-consistently solving the saddle-point equations as described above. The resulting phase diagram is shown in Fig.~\ref{fig2}. For attractive interactions between the two flavors ($\alpha < 0$) we find an SYK non-Fermi liquid with extensive residual entropy. In contrast, for repulsive interactions  ($\alpha > 0)$ there is an instability to a gapped superconducting phase generated by the spontaneous breaking of $U(1)$ charge conservation. This should be compared to the results of Refs.~\cite{Sahoo2020, Klebanov2020}, showing a spontaneous breaking of the \emph{axial} $U(1)$ symmetry with  quantum number $Q_- = Q_\up - Q_\down$, whereby an ``excitonic" order parameter $\frac{1}{N}\sum_j \langle c_{j \up} c_{j \down}^\dagger \rangle$ is generated for $\alpha < 0$. Indeed, the Hamiltonian studied in Refs.~\cite{Sahoo2020, Klebanov2020} is related to Eq.~\ref{eq:H_two_TR_cSYKS} by a particle-hole transformation for a single flavor $c_{i \down}^\dagger \leftrightarrow  c_{i \down}$  combined with $\alpha \rightarrow - \alpha$,
according to which we expect a spontaneous expectation value $\Delta \equiv F_{\tau=0} = \frac{1}{N} \sum_j \langle c_{j \up} c_{j \down} \rangle$ to develop for $\alpha > 0$. That is, in our case the \emph{global} U(1) symmetry with $\mathcal{Q} = \mathcal{Q}_\up + \mathcal{Q}_\down$ is instead broken, leading to a spin-singlet SC state, and the instability now interestingly occurs for \emph{repulsive} inter-flavor interactions.

\begin{figure}
\centering
\includegraphics[width=\columnwidth]{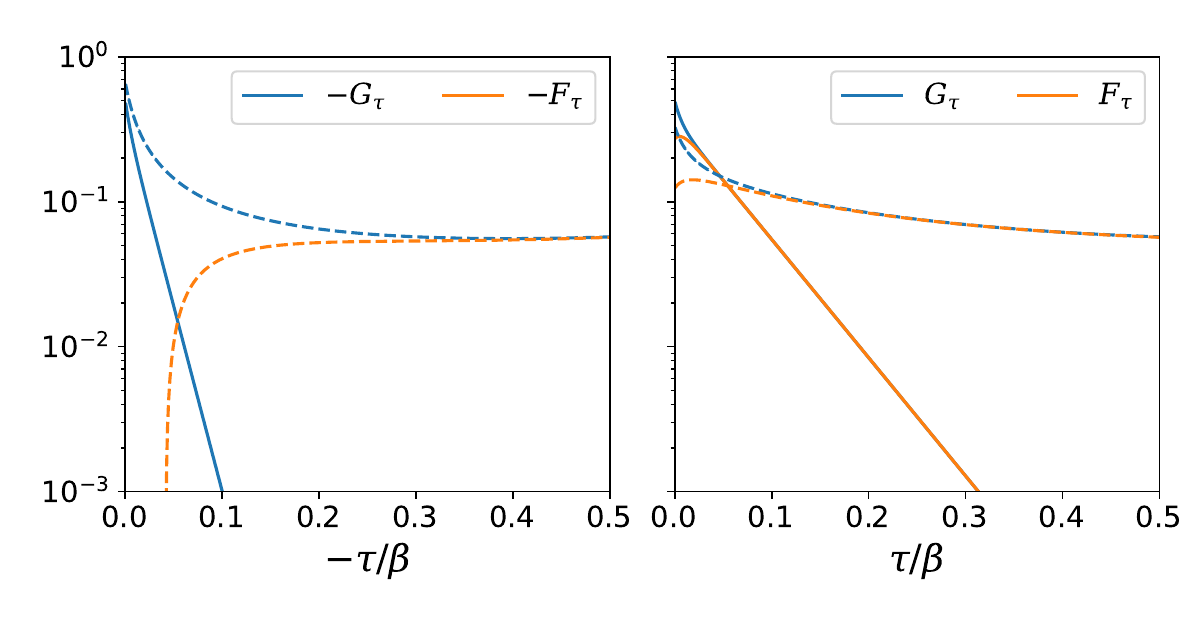}
    \caption{Comparison of the regular and anomalous Green's functions $G_\tau$ and $F_\tau$ in the gapped (solid lines, $\alpha=0.4$ and $B=0.1 J$) and gapless (dashed lines, $\alpha=0.4$ and $B=0.2J$) SC phases at low temperature $\beta J = 200$. We show both negative (left) and positive (right) imaginary times $\tau$.}
    \label{fig3}
\end{figure}

In the presence of a weak Zeeman field $B$, the SC phase remains non-magnetized ($M=0$) as expected for a fully gapped spin-singlet superconductor. The breaking of time-reversal symmetry is however reflected in the different spectral gaps for the hole and electron sides, as shown in Fig.~\ref{fig3}. In contrast, the non-Fermi liquid phase can be continuously magnetized by tuning $B$, a reflection of the compressibility of the underlying cSYK models~\cite{Sachdev2015, Gu2020}. At sufficiently large $B$ a first-order phase transition takes the system to a gapped, fully polarized state with $M = \frac{1}{2}$. The discontinuous jump in residual entropy between the non-Fermi liquid and gapped phases signals a first-order phase transition. The transition between the two gapped, ordered phases (SC with $\Delta \neq 0$ and polarized phase with $M = \frac{1}{2}$) is also of first order, as expected from standard Landau arguments. 

A surprising result is the appearance of an intermediate phase which is gapless and superconducting, upon applying a Zeeman field $B$. This phase exhibits extensive residual entropy and magnetization associated with the SYK non-Fermi liquid, as well as a non-zero SC order parameter $\Delta$. The presence of a non-zero $M$ and $\Delta$ seems contradictory, but can  occur e.g. in a ``phase coexistence" scenario where only part of the system spontaneously breaks the U(1) symmetry~\cite{Sahoo2020}. Here the Green's function $G_\tau$ exhibits power-law decay at long times, in contrast to the exponential decay observed in the gapped SC phase (see Fig.~\ref{fig3}). 
When tuning the chemical potential away from charge neutrality, $\mu \neq 0$, we find that both the gapped and gapless SC phases are compressible, as described in more detail in the Appendix.

\begin{figure*}
\includegraphics[width=\textwidth]{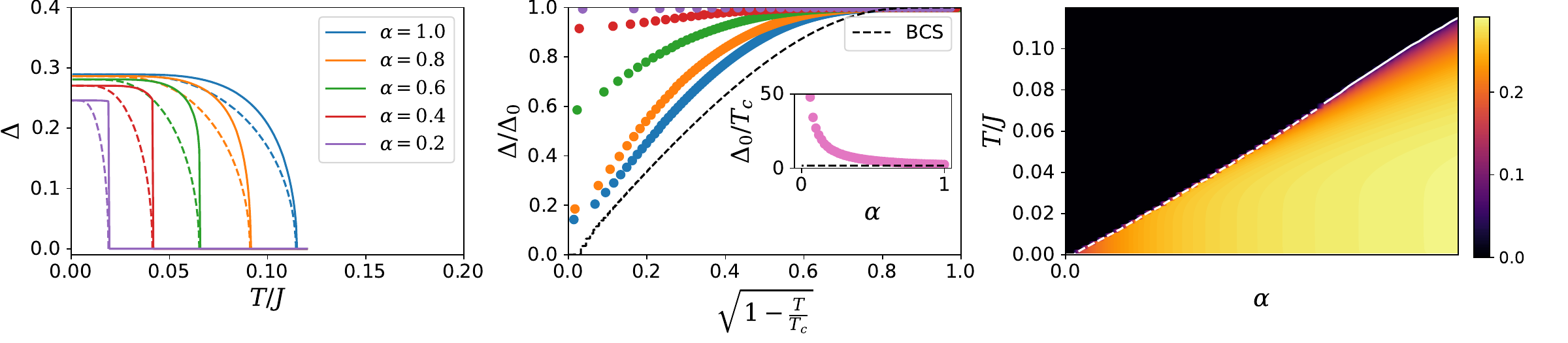}
    \caption{(Left:) Temperature dependence of the superconducting order parameter $\Delta$ for various values of $\alpha$ and $\mu=B=0$. The weak-coupling BCS scaling is shown by dashed lines. (Middle:) Data collapse of $\Delta/\Delta_0$ against $\sqrt{1 - T/T_c}$. There is a jump from a second to a first order phase transition when the interaction strength $\alpha$ decreases. (Inset:) The ratio $\Delta_0/T_c$ increases as $\alpha \rightarrow 0$ and is greatly enhanced compared to the BCS result (dashed line). (Right): Phase diagram showing $\Delta$ in the $T - \alpha$ plane, with second-order (solid line) and first-order (dashed line) phase transitions out of the gapped superconducting phase.
    \label{fig4}}
\end{figure*}

\emph{Gap ratio enhancement.--} We now increase temperature and consider the transition out of the gapped SC phase. In Fig.~\ref{fig4} we show the temperature dependence of $\Delta$ for $B=0$. For large $\alpha$ we find that $\Delta$ smoothly goes to zero at $T_c$, indicative of a second-order transition, which is however not BCS-like as shown from comparing with the self-consistent solution of the BCS gap equations in the weak coupling limit~\cite{Bardeen1957a, Bardeen1957b, Gorkovr1958}. In particular, in BCS theory the following universal relations hold (with $k_B=1$ and $\Delta_0$ the SC order parameter at $T=0$):
\begin{equation}
    \Delta_0 = 1.76 T_c ~~ , ~~ \Delta(T \rightarrow T_c) = 3.06 T_c \sqrt{1-\frac{T}{T_c}}.
\end{equation}
Here we find that neither relation is satisfied, highlighting the strongly-correlated nature of superconductivity. Further, the data collapse near $T_c$ suggests that the SC transition becomes of first order when decreasing $\alpha$. There is also a significant gap ratio enhancement~\cite{Patel2018a} with $\Delta_0/T_c$ seemingly diverging for small $\alpha$, which can be traced back to the empirical observation that $T_c \sim \alpha$ while $\Delta_0$ depends only weakly on the interaction strength. 

\emph{Equilibrium transport.--} We finally consider transport properties of the SC phases identified above. To do so we build a lattice model out of spinful SYK building blocks, connected by random hoppings similarly to Ref.~\cite{Song2017},
\begin{align}
H =& \sum_x H_x + \sum_{\langle x, x' \rangle} \sum_{ij \sigma} t^{x x'}_{ij \sigma} c^\dagger_{i \sigma x} c_{j \sigma x'} 
\label{eq:hopping_random_lattice}.
\end{align}
Here $H_x$ describe spinful SYK models, Eq.~\ref{eq:H_two_TR_cSYKS}, with an \emph{independent} disorder realization on each site $x$. This ensures that the effective action only features local Green's functions and self-energies. The hopping terms connect nearest neighbors $\langle x, x' \rangle $ and are drawn from a Gaussian distribution with zero mean and variance $\overline{| t_{ij \sigma}^{xx'}|^2} = \frac{t^2}{N}$.

To drive a supercurrent in the system we consider a ring geometry with $L$ sites threaded by a magnetic flux $\Phi$. This introduces Peierls phase factors in the hopping parameters through $t_{ij \sigma}^{xx'} e^{i \phi}$, with $\phi = \frac{e}{\hbar} \int  \mathbf{A} \cdot d\mathbf{l} = \frac{2 \pi}{L} \frac{\Phi}{\Phi_0}$ and the flux quantum $\Phi_0 = \frac{h}{e}$. If the hopping parameters are taken to be \emph{uncorrelated} between the two spin components, the disorder average yields only the Green's function $G_{x,\tau}$ which is insensitive to the magnetic flux insertion. It is thus crucial to require invariance of the hopping terms under the anti-unitary symmetry $\Theta$ -- that is, $t_{ij \up}^{xx'} = ( t_{ij \down}^{xx'})^*$. 
Combined with a translation-invariant ansatz, whereby $G_{x, \tau} = G_\tau$ and $F_{x, \tau}  = F_\tau$, we obtain saddle-point equations (see Appendix) that can be solved self-consistently. The free energy density $\mathcal{F}/L$ is computed using the appropriate generalization of Eq.~\ref{eq:free_energy}, with the induced supercurrent
\begin{equation}
I = \frac{ \partial \mathcal{F} }{\partial \Phi} =  \frac{2e}{\hbar} \frac{\partial}{\partial \varphi } \left( \frac{\mathcal{F}}{L} \right),
\label{eq:I_phi}
\end{equation} 
where $\varphi = 2 \phi$ is the phase carried by Cooper pairs when tunneling between SYK dots.

\begin{figure}
\centering
\includegraphics[width=\columnwidth]{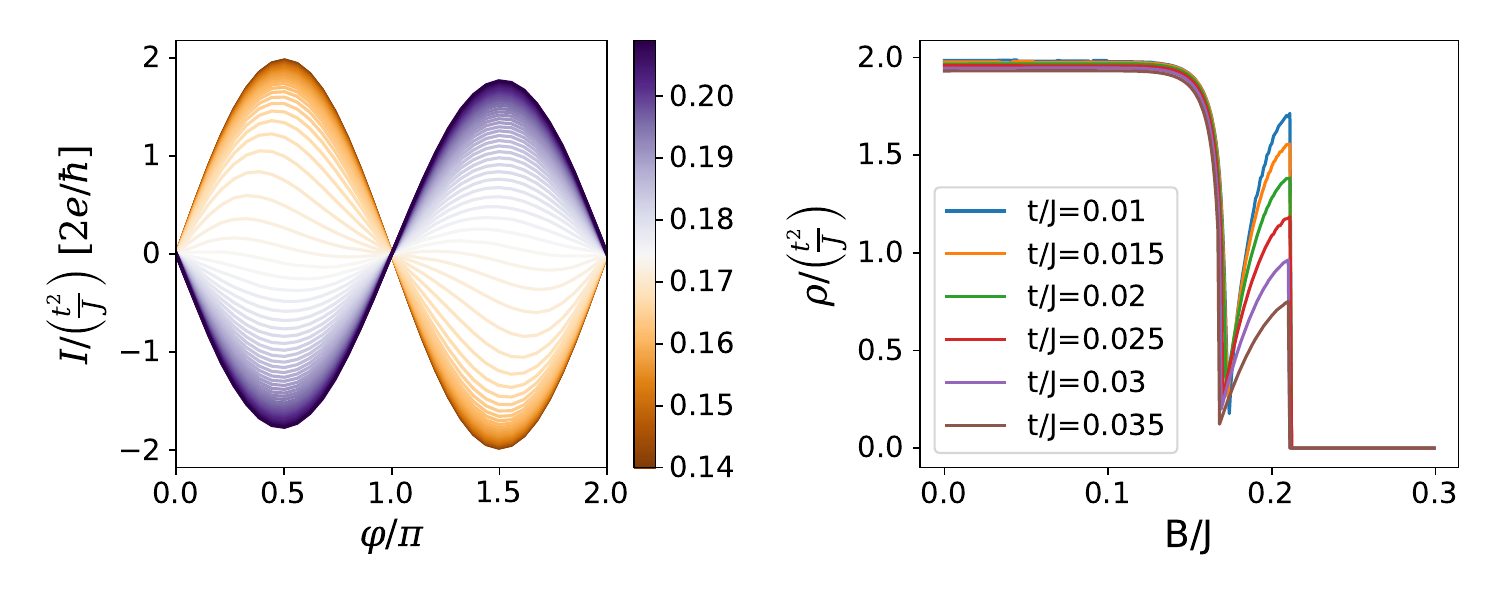}
    \caption[Supercurrent-phase relations in the lattice construction]{Equilibrium transport properties of the two superconducting phases, here for $\alpha=0.5$. (Left:) Supercurrent-phase relation $I(\varphi)$ [computed through Eq.~\ref{eq:I_phi}] in the lattice model for various values of the Zeeman field $B/J$ (color scale) and $t/J = 0.01$. The jump to a $\pi$-shifted sinusoidal profile coincides with the first-order transition between the gapped and gapless SC phases at the critical Zeeman field $B_c$. (Right:) The superfluid density $\rho$ (in arbitrary units) is independent of $B$ in the gapped phase and shows a recovery after a sudden drop at $B_c$.}
    \label{fig5}
\end{figure}

The limit of weak hopping $t$ corresponds to Josephson tunneling between neighboring SC islands that are phase biased. Accordingly, we obtain sinusoidal supercurrent-phase relations $I(\varphi) = I_c \sin \left( \varphi + \delta \right)$, as shown in Fig.~\ref{fig5} for $\alpha = 0.5$ and various values of $B$. In the gapped phase we find $\delta = 0$ and the maximal supercurrent $I_c \sim t^2 / J$, as expected in perturbation theory from the tunneling of Cooper pairs between neighboring sites. For sufficiently large $B$ the gapless SC phase is stabilized (see also Fig.~\ref{fig2}), which in transport is manifest as a phase-shifted supercurrent relation with $\delta = \pi$. In other words, the system's free energy is minimized for a staggered order parameter $\Delta_x$ with a $\pi$ phase difference between neighboring sites. The superfluid density $\rho \sim \frac{\partial I}{\partial \varphi} |_{\varphi \rightarrow \delta}$ is independent of $B$ in the gapped phase, but interestingly shows a recovery with $B$ in the gapless phase, following a sudden drop at the phase transition at $B_c$. The gapless SC phase is however more fragile to competing energy scales, as seen from the rapid decrease in $\rho/t^2$ as a function of $t$. 

\emph{Discussion.--} In this work we introduced a simple ``spinful SYK" model for strongly-correlated superconductivity. Its exact solvability in the large $N$ limit allowed us to map the model's phase diagram which exhibits \emph{two} different (gapped and gapless) superconducting phases, and show how their behavior strongly deviates from BCS theory. The transport properties of such phases, going beyond the equilibrium picture presented here, could be explored in future work. Indeed, the lattice model (Eq.~\ref{eq:hopping_random_lattice}) hosts not only correlated SC phases, but also a strange metal and a heavy Fermi liquid (depending on the ratio $t/J$) in the limit $\alpha=0$, where it reduces to two decoupled (spinless) SYK chains~\cite{Song2017}. It would be interesting to study the thermal and electrical conductivity across this rich phase diagram, which bears some resemblance to the phenomenology of cuprates.

In summary, this work adds to the growing body of literature on SYK superconductivity~\cite{Patel2018b, Esterlis2019, Cheipesh2019, Chowdhury2020a, Chowdhury2020b, Hauck2020, Wang2020a, Wang2020b, Wang2020c, Pan2021} by highlighting the role of anti-unitary symmetries in promoting SC instabilities. Further, the model's simple structure and connections to physical platforms where superconducting instabilities are expected for \emph{repulsive electronic interactions} raise the hope of stimulating new experimental developments. An interesting open question concerns the effect of (finite $N$) fluctuations away from the saddle-point, which should restore the broken U(1) symmetry at low energy in accordance with the Mermin-Wagner theorem~\cite{Maldacena2016, Bagrets2016}.

\emph{Acknowledgments.--} We are  grateful to Stephan Plugge for  discussions and collaborations on related projects, and to Nikolay Gnezdilov and Yuxuan Wang for insightful comments on the first version of this manuscript. This research was supported in part by the National Sciences and Engineering Council of Canada (NSERC) and the Canada First Research Excellence Fund (CFREF).

\bibliography{ref}

\appendix

\clearpage

\begin{widetext}

\section{Experimental platforms and electronic interactions}
\label{app:exp_interactions}

In this Appendix we consider a simple solid-state platform that provides an approximate physical realization of the spinful SYK model, Eq.~\ref{eq:H_two_TR_cSYKS}. This platform is largely inspired by Refs.~\cite{Chen2018, Sahoo2020}, but the same symmetry setting might be relevant for other platforms based on ultracold atoms~\cite{Danshita2017}, optical lattices~\cite{Wei2021} or spin chains~\cite{Zhou2020}.

We consider a (0+1)-dimensional ``quantum dot" geometry inspired by the graphene flake of Ref.~\cite{Chen2018}. Here we promote this setup to a topological insulator (TI) flake with the two surfaces denoted by $1,2$. Each surface hosts a single Dirac fermion, which can described by the low-energy Hamiltonians
\begin{align}
    h_1( {\bm k} ) = + \hbar v_F {\bm \sigma} \cdot {\bm k} - \mu_1 \mathds{I} \quad , \quad
    h_2( {\bm k} ) = - \hbar v_F {\bm \sigma} \cdot {\bm k} - \mu_2 \mathds{I} ,
\label{eq:TI_flake_hk}
\end{align}
where ${\bm \sigma}$ are Pauli matrices acting on the electron spin, $v_F$ is the Fermi velocity and $\mu_{1/2}$ are chemical potentials, which in general could be different on the two surfaces (e.g. due to the TI flake being deposited on a substrate). The opposite signs of the Fermi velocity in Eq.~\ref{eq:TI_flake_hk} capture the fact that the two TI surfaces have opposite normal vectors. 

When a strong (perpendicular) magnetic field $\bm{B} = \nabla \times \bm{A} = B \bm{\hat{z}}$ is applied to the sample (see Fig.~\ref{fig6}), the Dirac surface states collapse to a series of flat Landau levels~\cite{Cheng2010,Yoshimi2015}.
The low-energy theory for the two surfaces reads
\begin{align}
    h_1( {\bm k} ) = +  \hbar v_F {\bm \sigma} \cdot \left( {\bm k} + e {\bm A} \right) + B_Z \sigma^z - \mu_1 \mathds{I} , \nonumber \\
    h_2( {\bm k} ) = -  \hbar v_F {\bm \sigma} \cdot \left( {\bm k} + e {\bm A} \right) + B_Z \sigma^z - \mu_2 \mathds{I},
    \label{eq:surfacestates_uniformB}
\end{align}
with the Zeeman energy $B_Z = \frac{1}{2} g \mu_B B$ where $g$ is the Land\'e factor and $\mu_B$ the Bohr magneton. For $\mu_1 = \mu_2$ the two surfaces are related by the unitary rotation $\sigma^z$ -- in other words the system is invariant under the unitary $U = \tau^x \sigma^z$ where $\tau^x$ is a Pauli matrix acting on the surface pseudospin. Let us set the chemical potential of each surface to lie within its respective zeroth Landau level  at energy $E_0 = -B_Z$. Similarly to the case of graphene, where the zeroth Landau level is sublattice polarized (within each valley), the zeroth Landau level of a TI flake is spin-polarized, with
\begin{equation}
    \phi_j(\mathbf{r}) = 
    \begin{pmatrix}
        0 \\
        \phi_{j1\down}( \mathbf{r} ) \\
        0\\
        \phi_{j2\down}( \mathbf{r} )
    \end{pmatrix}.
\end{equation}
Here $j$ labels the degenerate LL$_0$ wavefunctions, and the unitary symmetry $U$ imposes the constraint $\phi_{j1\downarrow}( \mathbf{r} ) = \phi_{j2\downarrow}( \mathbf{r} )$. Following the reasoning described in Ref.~\cite{Chen2018, Sahoo2020}, when including Coulomb interactions within a strongly disordered zeroth-Landau level manifold (where the disorder mainly comes from a controllable source such as the irregular boundary of the flake, such that it is correlated between the two surfaces), this setup leads to an approximate physical realization of coupled identical cSYK models, as analyzed in  Ref.~\cite{Sahoo2020, Klebanov2020}. (More precisely, each surface would be described by a sparse or ``low-rank" SYK model which nevertheless shows interesting conformal behavior~\cite{Kim2020}).

One can imagine a slightly different setup where an effective time-reversal symmetry is preserved globally, while time reversal is broken at the level of an individual surface. This could be accomplished by using an inhomogeneous field configuration that points mostly towards (or away from) the TI flake, using e.g. two bar magnets with their north poles pointing towards the flake as shown in Fig.~\ref{fig6}. In this hypothetical setup the magnetic field on the two surfaces is opposite, with
\begin{align}
    h_1( {\bm k} ) &= +  \hbar v_F {\bm \sigma} \cdot \left( {\bm k} + e {\bm A} \right) + B_Z \sigma^z - \mu_1 \mathds{I} 
    \nonumber \\
    h_2( {\bm k} ) &= -  \hbar v_F {\bm \sigma} \cdot \left( {\bm k} - e {\bm A} \right) -  B_Z \sigma^z - \mu_2 \mathds{I} .
    \label{eq:TI_flake_TRS}
\end{align}
When $\mu_1 = \mu_2$ the two surface theories are now time-reversed partners, with the anti-unitary time-reversal operator taking the form $\Theta = \tau^x \sigma^x \mathcal{K}$ with $\Theta ^2= +1$. This can be understood as a combination of the spinful TRS $i \sigma^y \mathcal{K}$ with the unitary rotation $U = \tau^x \sigma^z$ mentioned above. As a consequence, the spin polarization of LL$_0$ wavefunctions (again with energy $E_0 = -B_Z$) is opposite on the two surfaces, with
\begin{equation}
    \phi_j(\mathbf{r}) = 
    \begin{pmatrix}
        0 \\
        \phi_{j1\down}( \mathbf{r} ) \\
        \phi_{j2\up}( \mathbf{r} ) \\
        0 \\
    \end{pmatrix},
\end{equation}
and time-reversal $\Theta$ imposes the constraints $\phi_{j 1 \down}(\mathbf{r}) = \phi_{j 2 \up}^*(\mathbf{r})$. The surface and spin degrees of freedom being locked, we use a single index $a = \up, \down$ in the following to denote the LL$_0$ states on the two surfaces, matching the notation in the main text. 
\begin{figure}
\centering
\includegraphics[width=0.9\columnwidth]{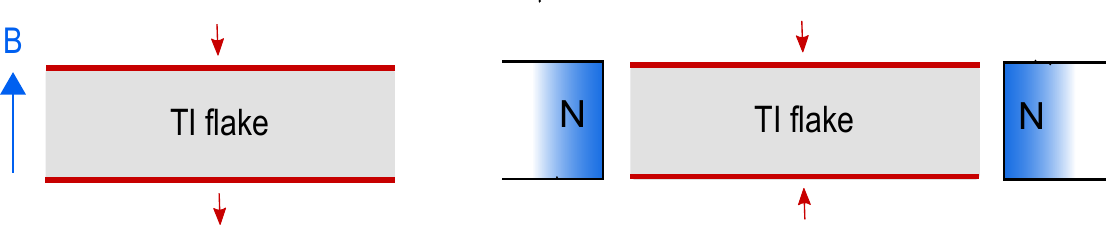}
    \caption{Schematic diagram of proposed physical realizations of coupled cSYK models using the lowest Landau level (LL$_0$) surface states (marked in red, with the corresponding spin polarization denoted by a red arrow) of a topological insulator (TI) flake with an irregular boundary. (Left): Under a strong perpendicular magnetic field $\bm{B}$ the two surfaces host identical spin-polarized LL$_0$ states, leading to an approximate physical realization of the model discussed in Refs.~\cite{Sahoo2020, Klebanov2020}, where an instability to an excitonic phase is expected for attractive inter-flavor interactions. (Right): A pair of bar magnets creates a magnetic field configuration that points in opposite directions at the two surfaces, leading to time-reversed LL$_0$ states with opposite spin projections. This provides an approximate physical realization of the spinful SYK model in Eq.~\ref{eq:H_two_TR_cSYKS}, where superconducting instabilities are expected for repulsive inter-flavor interactions.}
    \label{fig6}
\end{figure}

Let us now analyze the form of interactions. Within the LL$_0$ manifold, we consider the (projected) Coulomb interactions
\begin{align} 
H_{C} = \frac{1}{2} \sum_{a, b} \sum_{\mathbf{r},\mathbf{r}'} \rho_{a}(\mathbf{r}) V_{a b} (\mathbf{r}-\mathbf{r}') \rho_{b}(\mathbf{r}')
\end{align}
where $V_{a b}(\mathbf{r}-\mathbf{r'})$ is the screened Coulomb potential. In the graphene flake setup of Ref.~\cite{Chen2018}, where $a$ and $b$ denote spin projections living \emph{in the same spatial region}, the Coulomb potential does not distinguish between spin components, leading to SU(2) symmetric interactions~\cite{Sahoo2020}. However, in the TI flake where $a,b = \up, \down$ denote spin-polarized LL$_0$ states on different surfaces, the intra-surface interactions are expected to be stronger than inter-surface ones. We thus set $V_{\up \up} (\mathbf{r}) = V_{\down \down} (\mathbf{r}) = V_0 (\mathbf{r})$ and $V_{\up \down} (\mathbf{r}) = V_{\down \up}(\mathbf{r}) = \alpha V_0 (\mathbf{r})$ with  $0 < \alpha < 1$ expected for repulsive Coulomb interactions. We however consider both signs of $\alpha$ in the main text for completeness.

The local charge density at point $\mathbf{r}$ reads
\begin{align}
    \rho_{a}(\mathbf{r}) = c^\dagger_{\mathbf{r} a} c_{\mathbf{r} a} = \sum_{ik} \phi_{i a}^*(\mathbf{r}) \phi_{k a}(\mathbf{r}) c_{i a}^\dagger c_{k a}
\end{align}
in terms of the LL$_0$ wavefunctions, which leads to an interaction Hamiltonian $H_\alpha = \sum_{a,b} \sum_{ijkl} J^{a b}_{ij;kl} c_{i a}^\dagger c_{j b}^\dagger c_{k a} c_{l b} $ with interaction parameters
\begin{align} 
 J^{a b}_{ij;kl} = - \frac{1}{2} \sum_{\mathbf{r}, \mathbf{r'}} \phi_{i a}^*(\mathbf{r}) \phi_{j b}^*(\mathbf{r'})  V(\mathbf{r} - \mathbf{r'}) \phi_{k a}(\mathbf{r})  \phi_{l b}(\mathbf{r'}),
\end{align}
where we assumed that all indices $i,j,k,l$ are different. This is a useful simplifying assumption also used e.g. in Refs.~\cite{Chen2018,Gu2020,Sahoo2020} (see also footnote [41]).
The anti-symmetry $\Theta$ imposes the following symmetries on the tensor of couplings:
\begin{align} 
 J^{\down \down}_{kl;ij} &= J^{\up \up}_{ij;kl} \quad , \quad  J^{\up \down}_{il;kj} = \alpha J^{\up \up}_{ij;kl}, \nonumber \\
 J^{\down \up}_{kl;ij} &= J^{\up \down}_{ij;kl} \quad , \quad
 J^{\down \up}_{kj;il} = \alpha J^{\up \up}_{ij;kl}
\label{eq:constraints}.
\end{align}
Because Hermiticity imposes $ J^{a b}_{ij;kl} = \left( J^{a b}_{kl;ij}\right)^*$, we have $ J^{\down \down}_{ij;kl} = \left( J^{\up \up}_{ij;kl}\right)^*$ --
the two cSYK models have coupling constants that are complex conjugated. The Hamiltonian thus reads, using the constraints in Eq.~\ref{eq:constraints},
\begin{align}
H_\alpha =& \sum_{ijkl} J_{ij;kl} \Big[ c_{i \up}^\dagger c_{j \up}^\dagger c_{k \up} c_{l \up} +  c_{k \down}^\dagger c_{l \down}^\dagger c_{i \down} c_{j \down} 
 + \alpha \left( c_{i \up}^\dagger c_{l \down}^\dagger c_{k \up} c_{j \down} + c_{k \down}^\dagger c_{j \up}^\dagger c_{i \down} c_{l \up} \right)  \Big],
\end{align}
as in Eq.~\ref{eq:H_two_TR_cSYKS} in the main text.

\section{Effective action}
\label{app:saddle_point}

We now derive the large-N saddle-point equations of our model, mostly following Ref.~\cite{Sahoo2020}. We start by writing the corresponding partition function in the Euclidean time formalism at inverse temperature $\beta = 1/T$,
\begin{align} 
\mathcal{Z} = \int \mathcal{D}[c, c^\dagger]  \exp\left[ - \int_0^\beta d\tau \left( \sum_{i,a=\up, \down} c^{\dagger}_{i a}(\tau) \partial_\tau  c_{i a}(\tau)  +  H \right) \right] , \end{align}
where $H$ is given in Eq.~\ref{eq:H_two_TR_cSYKS}. We first rewrite the interaction terms using only independent coupling constants by restricting the sum,
\begin{align}
H_\alpha =& \sum_{i<j,k<l} J_{ij;kl} \Big[ 4 \left( c_{i \up}^\dagger c_{j \up}^\dagger c_{k \up} c_{l \up} +  c_{k \down}^\dagger c_{l \down}^\dagger c_{i \down} c_{j \down} \right) + 2 \alpha \left( c_{i \up}^\dagger c_{l \down}^\dagger c_{k \up} c_{j \down} - c_{j \up}^\dagger c_{l \down}^\dagger c_{k \up} c_{i \down} - c_{i \up}^\dagger c_{k \down}^\dagger c_{l \up} c_{j \down} + c_{j \up}^\dagger c_{k \down}^\dagger c_{l \up} c_{i \down} 
 \right) \Big].
\end{align}
We then obtain the partition function corresponding to the quenched average over the Gaussian-distributed coupling constants,
$ \mathcal{Z}_{\rm avg} = \int d[J, J^*]  P(J_{ijkl}) \mathcal{Z} = \int D[c, c^\dagger] e^{-S}$ with $P(J_{ijkl}) = {\rm exp} \left( - \frac{ |J_{ijkl}|^2 }{\sigma^2} \right) $, which leads to the effective action 
\begin{align}
S =  \int_0^\beta d\tau \left( \sum_{i,a} c^{\dagger}_{i a}(\tau) \left( \partial_\tau - \mu_a \right) c_{i a}(\tau)  \right) -  \frac{\sigma^2}{2}  \sum_{i<j;k<l} \phi_{ijkl} \phi_{klij},
\label{eq:effective_action}
\end{align} 
 with the variance $\sigma^2 \equiv \overline{ | J_{ijkl} |^2} = J^2/8N^3$, and where we defined
\begin{align}
    \phi_{ijkl} = 4 \int_0^\beta d\tau \left[\left( c_{i \up}^\dagger c_{j \up}^\dagger c_{k \up} c_{l \up} +  c_{k \down}^\dagger c_{l \down}^\dagger c_{i \down} c_{j \down} \right) + \frac{\alpha}{2} \left( c_{i \up}^\dagger c_{l \down}^\dagger c_{k \up} c_{j \down} - c_{j \up}^\dagger c_{l \down}^\dagger c_{k \up} c_{i \down} -
    c_{i \up}^\dagger c_{k \down}^\dagger c_{l \up} c_{j \down} + c_{j \up}^\dagger c_{k \down}^\dagger c_{l \up} c_{i \down} 
 \right) \right],
\end{align}
with the imaginary time dependence of the Grassmann fields implied. Expanding this term and replacing $\sum_{i<j;k<l} \rightarrow \frac{1}{4} \sum_{ijkl}$, the second term in Eq.~\ref{eq:effective_action} becomes
\begin{align}
    \frac{-J^2}{4  N^3} \sum_{ijkl} &\int_0^\beta d\tau \left[ c_{i \up}^\dagger c_{j \up}^\dagger c_{k \up} c_{l \up} +  c_{k \down}^\dagger c_{l \down}^\dagger c_{i \down} c_{j \down} + \frac{\alpha}{2} \left( c_{i \up}^\dagger c_{l \down}^\dagger c_{k \up} c_{j \down} - c_{j \up}^\dagger c_{l \down}^\dagger c_{k \up} c_{i \down} - c_{i \up}^\dagger c_{k \down}^\dagger c_{l \up} c_{j \down} + c_{j \up}^\dagger c_{k \down}^\dagger c_{l \up} c_{i \down} 
 \right) \right] \nonumber \\
    \times &\int_0^\beta d\tau' \left[ c_{k \up}^\dagger c_{l \up}^\dagger c_{i \up} c_{j \up} +  c_{i \down}^\dagger c_{j \down}^\dagger c_{k \down} c_{l \down} + \frac{\alpha}{2} \left( c_{k \up}^\dagger c_{j \down}^\dagger c_{i \up} c_{l \down} - c_{k \up}^\dagger c_{i \down}^\dagger c_{j \up} c_{l \down} - c_{l \up}^\dagger c_{j \down}^\dagger c_{i \up} c_{k \down} + c_{l \up}^\dagger c_{i \down}^\dagger c_{j \up} c_{k \down} 
 \right) \right].
\end{align}
Expanding this product, one sees the fundamental difference with the solution of coupled cSYK models in the presence of a unitary symmetry~\cite{Sahoo2020, Klebanov2020}: the disorder average yields anomalous terms of the form $\sum_i c_{i \up}^\dagger c^\dagger_{i \down}$. These terms will lead to spontaneous breaking of the global U(1) symmetry if they develop a finite expectation value in the saddle-point solutions.

In order to integrate the fermion fields we introduce Green's functions and their associated self-energies using
\begin{align}
1 &\sim \int \mathcal{D}\Sigma_\up \mathcal{D}G_\up \exp  \left( N \int d\tau d\tau' \Sigma_\up(\tau, \tau') \left [G_\up(\tau', \tau) - \frac{1}{N}\sum_{i=1}^{N} c_{i \up}(\tau') c^{\dagger}_{i \up}(\tau) \right] \right), \nonumber \\
1 &\sim \int \mathcal{D}\Sigma_\down \mathcal{D}G_\down \exp  \left( N \int d\tau d\tau' \Sigma_\down(\tau, \tau') \left [G_\down(\tau', \tau) - \frac{1}{N}\sum_{i=1}^{N} c^\dagger_{i \down}(\tau') c_{i \down}(\tau) \right] \right),
\label{eq:identity}
\end{align}
as well as their anomalous counterparts
\begin{align}
1 &\sim \int \mathcal{D}\Pi \mathcal{D} F \exp  \left( N \int d\tau d\tau' \Pi(\tau, \tau') \left [F(\tau', \tau) - \frac{1}{N}\sum_{i=1}^{N} c_{i \up }(\tau') c_{i \down}(\tau) \right] \right) ,\nonumber \\
1 &\sim \int \mathcal{D}\tilde{\Pi} \mathcal{D} \tilde{F} \exp  \left( N \int d\tau d\tau' \tilde{\Pi}(\tau, \tau') \left [\tilde{F}(\tau', \tau) - \frac{1}{N}\sum_{i=1}^{N} c^\dagger_{i \down}(\tau') c^\dagger_{i \up}(\tau) \right] \right) .
\label{eq:identity_anomalous}
\end{align}
Exploiting time translation invariance, whereby $ G_a(\tau,\tau') = G_a(\tau-\tau')$ and so on, we get
\begin{align}
S &=  S_0 - N \beta  \int_0^\beta d\tau \Big[ \sum_a  \Sigma_{a}(-\tau) G_{a}(\tau) + \Pi(-\tau) F(\tau) + \tilde{\Pi}(-\tau) \tilde{F}(\tau) \nonumber\\
&+ \frac{J^2}{4}  \Big\{ \sum_a G_{a}^2(\tau) G_{a}^2(-\tau) + 2 F^2(\tau) \tilde{F}^2(-\tau) - 4 \alpha \sum_a G_a(\tau) G_a(-\tau) F(\tau) \tilde{F}(-\tau)  \nonumber \\
    & + \alpha^2 \Big( G_{\up}(\tau) G_{\down}(\tau) G_{\up}(-\tau)  G_{\down}(-\tau) + F(-\tau) F(\tau) \tilde{F}(\tau) \tilde{F}(-\tau) 
 + 2 G_{\up}(\tau) G_{\down}(\tau)  F(-\tau) \tilde{F}(-\tau)  \Big)    \Big\} \Big].
\label{eq:GSigma_Action}
\end{align} 
Here $S_0$ denotes the free fermion part of the action, which must be analyzed in Nambu space to account for the anomalous pairing terms generated by the SYK interactions. Writing $\mu_{\up, \down} = \mu \pm B$ as in Eq.~\ref{eq:H_two_TR_cSYKS}, we have
\begin{equation}
    S_0 = \sum_j  \int d\tau d\tau' \Psi_j^\dagger(\tau) \left[ A \delta(\tau-\tau') + \partial_\tau \delta_{ab} \delta(\tau-\tau') - \Sigma(\tau, \tau') \right] \Psi_j(\tau') ,
\end{equation}
with the Nambu spinors $\Psi_j = ( c_{j \up}, c_{j \down}^\dagger )^T$ and the matrices
\begin{equation}
 A =   \begin{pmatrix}
    -\mu - B & 0 \\
    0  & \mu - B \\
    \end{pmatrix}  \quad , \quad
    \Sigma(\tau,\tau') =   \begin{pmatrix} 
    \Sigma_{\up}(\tau, \tau') & \tilde{\Pi}(\tau, \tau') \\
    \Pi(\tau, \tau') & \Sigma_{\down}(\tau, \tau')  \\
    \end{pmatrix}.
\end{equation}
Using time translation invariance to express $\Sigma(\tau,\tau') = \Sigma(\tau-\tau')$, we Fourier transform the action in terms of Matsubara frequencies $\omega_{n} = {(2n+1)\pi/\beta}$. Integrating out the Grassmann fields using the spinors $\Psi_n = ( c_{j\up}(\omega_n), c_{j \down}^\dagger(-\omega_n) )^T$, we thus obtain $S_0 = -N \ln \det M $ with
\begin{equation}
  M=  \bigoplus_n \begin{pmatrix} 
      -\mu - B -\Sigma_\up(\omega_n) - i\omega_n  & - \tilde{\Pi}(\omega_n) \\
 - \Pi(\omega_n) & \mu -B - \Sigma_\down(\omega_n) - i \omega_n  \\
    \end{pmatrix} .
\end{equation}

\subsection{Saddle-point equations}

Putting everything together, the action now reads 
\begin{align}
    -\frac{ S }{N} &= \ln \det M  + \sum_{\omega_n} \left( \Sigma_\up(\omega_n) G_\up(\omega_n) + \Sigma_\down(\omega_n) G_\down(\omega_n) +  \Pi(\omega_n) F(\omega_n) +  \tilde{\Pi}(\omega_n) \tilde{F}(\omega_n) \right) \nonumber \\
    &+ \frac{\beta J^2}{4} \int_0^\beta d\tau \Big\{ \sum_a G_a^2(\tau) G_a^2(-\tau) + 2 F^2(\tau) \tilde{F}^2(-\tau) - 4 \alpha \sum_a G_a(\tau) G_a(-\tau) F(\tau) \tilde{F}(-\tau)  \nonumber \\
    & + \alpha^2 \Big( G_{\up}(\tau) G_{\down}(\tau) G_{\up}(-\tau)  G_{\down}(-\tau) + F(-\tau) F(\tau) \tilde{F}(-\tau) \tilde{F}(\tau) 
   + 2 G_{\up}(\tau) G_{\down}(\tau) F(-\tau) \tilde{F}(-\tau) \Big)    \Big\} .
\end{align}
We obtain the saddle-point equations by taking functional derivatives of the action:
\begin{align}
\Sigma_{\up}(\tau) &= - J^2 \left[G^2_\up(\tau) G_{\up}(-\tau) -  2 \alpha G_\up(\tau) F(-\tau)  \tilde{F}(\tau) + \frac{\alpha^2}{2}\left( G_\up(\tau) G_\down(\tau) G_\down(-\tau) +  G_\down(-\tau) F(\tau) \tilde{F}(\tau) \right)  \right], \nonumber \\
\Sigma_{\down}(\tau) &=  -J^2 \left[G^2_\down(\tau) G_{\down}(-\tau)  -  2 \alpha G_{\down}(\tau)  F(\tau)  \tilde{F}(-\tau) + \frac{\alpha^2}{2}\left( G_\down(\tau) G_\up(\tau) G_\up(-\tau) +   G_\up(-\tau) F(\tau) \tilde{F}(\tau) \right)  \right], \nonumber \\
\Pi(\tau) &= -J^2 \left[ \tilde{F}^2(\tau) F(-\tau) - \alpha \tilde{F}(\tau) \sum_a G_a(\tau) G_a(-\tau)  +  \frac{\alpha^2}{2} \left( F(\tau) \tilde{F}(\tau) \tilde{F}(-\tau) +  G_\up(\tau) G_\down(\tau) \tilde{F}(-\tau)  \right)\right] ,\nonumber \\
\tilde{\Pi}(\tau) &= -J^2 \left[ F^2(\tau) \tilde{F}(-\tau) -  \alpha F(\tau) \sum_a G_a(\tau) G_a(-\tau) +  \frac{\alpha^2}{2} \left( \tilde{F}(\tau) F(\tau) F(-\tau) +  G_\up(\tau) G_\down(\tau) F(-\tau) \right) \right] ,
\label{eq:SD_equations_tau}
\end{align}
and 
\begin{align}
G_\up(\omega_n) &= \frac{\mu - B - \Sigma_\down(\omega_n) - i \omega_n }{D(\omega_n)}   \quad , \quad
 G_\down (\omega_n) = - \frac{\mu + B + \Sigma_\up(\omega_n) + i \omega_n }{D(\omega_n)}  , \nonumber \\
F(\omega_n) &=\frac{ \tilde{\Pi}(\omega_n)}{D(\omega_n)} \quad , \quad
\tilde{F}(\omega_n) = \frac{ \Pi(\omega_n)}{D(\omega_n)} , \nonumber\\
D(\omega_n) &=  (\mu + B + \Sigma_\up(\omega_n) + i \omega_n)(-\mu + B + \Sigma_\down(\omega_n) + i \omega_n ) - \Pi(\omega_n) \tilde{\Pi}(\omega_n) .
\label{eq:SD_equations_omega}
\end{align}
These are the general saddle-point equations, valid without additional symmetries, and are used to analyze the model in Eq.~\ref{eq:H_two_TR_cSYKS} with $\mu \neq 0$. At charge neutrality ($\mu=0$), the Green's functions respect $G_\up(\tau) = G_\down(\tau)$ and $F(\tau) = \tilde{F}(\tau)$ (and similarly for the self-energies). The saddle-point equations above thus simplify to
\begin{align}
\Sigma(\tau) &= - J^2 \left[ \left(1 + \frac{\alpha^2}{2} \right) G^2(\tau) G(-\tau) - 2 \alpha G(\tau) F(\tau) F(-\tau) + \frac{\alpha^2}{2} F^2(\tau) G(-\tau)  \right] , \nonumber \\
\Pi(\tau) &= -J^2 \left[ \left(1 + \frac{\alpha^2}{2} \right) F^2(\tau) F(-\tau) - 2 \alpha F(\tau) G(\tau) G(-\tau) + \frac{\alpha^2}{2} G^2(\tau) F(-\tau)  \right],
\end{align}
\begin{align}
G(\omega_n) &= - \frac{B + \Sigma(\omega_n) + i \omega_n }{D(\omega_n)}   \quad , \quad
F(\omega_n) = \frac{\Pi(\omega_n)}{D(\omega_n)} , \nonumber\\
D(\omega_n) &=  (B + \Sigma(\omega_n) + i \omega_n)^2- \Pi^2(\omega_n),
\end{align}
as in Eqs.~(\ref{eq:saddle_point_equations}) in the main text.

\section{Effective action for the lattice model}
\label{app:saddlepoint_lattice}

We now extend the saddle-point calculation to the lattice construction discussed in the main text, focusing on the charge neutrality point ($\mu=0$) for simplicity. The lattice model, Eq.~\ref{eq:hopping_random_lattice} contains two independent sets of Gaussian-distributed random variables (the SYK couplings $J_{ijkl}$ and the hopping parameters $t_{ij}$), which can be averaged separately. The previous solution for the spinful SYK model thus carries over, with an additional lattice site index $x$. We have
\begin{equation}
    \mathcal{Z}_{avg} = \int d[J,J^*] d[t,t^*] P(t_{ij}) P(J_{ijkl}) \mathcal{Z} = \int D[G, \Sigma] e^{- \left(S_l + S_t\right)} ,
\end{equation}
with the ``local" action $S_l$ as defined above, but with an explicit $x$ dependence in the correlators and self energies,
\begin{align}
    \frac{ S_l }{N} &= - \sum_x \ln \det M_x  - 2 \sum_{n,x} \left( \Sigma_x(\omega_n) G_x(\omega_n) +  \Pi_x(\omega_n) F_x(\omega_n) \right) \\
    &- \frac{\beta J^2}{2} \sum_x \int_0^\beta d\tau \Bigg\{ \left(1 + \frac{\alpha^2}{2}\right) \left( G_x^2(\tau) G_x^2(-\tau) + F_x^2(\tau) F_x^2(-\tau) \right) - 4 \alpha G_x(\tau) G_x(-\tau) F_x(\tau) F_x(-\tau)
     + \alpha^2 G_x^2(\tau) F_x^2(-\tau) \Bigg\}, \nonumber 
\end{align}
as well as the hopping contribution $S_t$ which, mirroring the steps leading to Eq.~\ref{eq:effective_action}, reads
\begin{align}
    S_t &= -  \frac{t^2}{N} \sum_{i,j} \sum_{\langle x, x' \rangle} | A_{ij}^{xx'} |^2  \quad , \quad
    A_{ij}^{xx'} = \int_0^\beta d\tau \left( e^{i \phi} c_{i \up x}^\dagger c_{j \up x'} + e^{-i \phi} c_{j \down x'}^\dagger c_{i \down x} \right),
\end{align}
with the phase $\phi = \frac{2 \pi}{L} \frac{\Phi}{\Phi_0}$ as introduced in the main text. Expanding this expression, using translation invariance in imaginary time and the Green's function definitions in Eqs.~\ref{eq:identity} and \ref{eq:identity_anomalous} leads to
\begin{align}
    \frac{ S_t }{N} &= 2 \beta t^2 \sum_{\langle x,x' \rangle} \int d\tau  \left\{ G_{x'}(\tau) G_{x}(-\tau) - F_{x'}(\tau) F_x(-\tau) \cos 2\phi \right\}.
    \label{eq:action_hopping_averaged}
\end{align}
 As remarked in the main text, we stress that the presence of the anomalous correlators $F(\tau)$ in the hopping contribution to the effective action, Eq.~(\ref{eq:action_hopping_averaged}) relies on the time-reversal symmetry in the hopping parameter distribution. This is crucial in obtaining a finite supercurrent, as the magnetic flux $\Phi$ only renormalizes the anomalous correlators.

Taking functional derivatives of this effective action then leads to the modified self-energy equations
\begin{align}
\Sigma_\tau &= -J^2 \Big[ \left( 1+\frac{\alpha^2}{2} \right) G^2_\tau G_{-\tau} - 2 \alpha G_{\tau} F_{\tau} F_{-\tau} + \frac{\alpha^2}{2}  F^2_{\tau} G_{-\tau} \Big] + z t^2 G_\tau \nonumber \\
\Pi_{\tau} &= -J^2 \Big[ \left( 1+\frac{\alpha^2}{2} \right) F^2_{\tau} F_{-\tau} - 2 \alpha  F_{\tau} G_{\tau} G_{-\tau} + \frac{\alpha^2}{2} G^2_{\tau}  F_{-\tau} \Big] - z t^2 \cos(2\phi) F_\tau .
\label{eq:saddle_point_equations_lattice}
\end{align}
Here we assumed that the averaged correlators respect translation symmetry at the saddle-point level, with the ansatz $G_x(\tau) = G(\tau)$, $F_x(\tau) = F(\tau)$ and similarly for the self-energies. The sum over nearest neighbors then simply contributes a factor of $z$ which is the coordination number of the lattice ($z=2$ for our ring model threaded by a magnetic flux), while the remaining sums over $x$ become trivial and yield a factor of $L$ (the number of lattice sites). The free energy density can be computed by substituting the saddle-point equations in the effective action,
\begin{align} 
\frac{\mathcal{F}}{L} &= -T \left[ 2 \ln 2 + \sum_n  \ln \left( \frac{D_n}{(i \omega_n)^2} \right) +\frac{3}{2} \sum_n \left\{ \Sigma_n G_n +  \Pi_n F_n  - z t^2 \left(G_n^2 - F_n^2 \cos 2 \phi \right) \right\} \right].
\label{eq:free_energy_complex_lattice}
\end{align} 

\end{widetext}

\section{Finite chemical potential}

\begin{figure}
\centering
\includegraphics[width=\columnwidth]{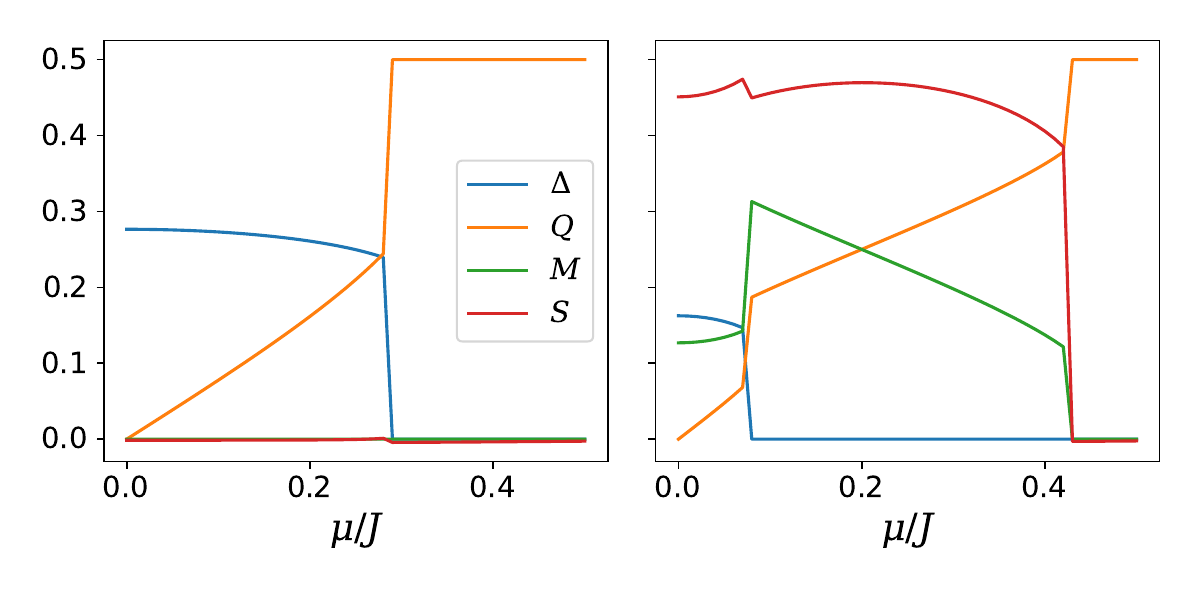}
    \caption{Stability of superconductivity to tuning the chemical potential $\mu$ away from charge neutrality. We take $\alpha=0.5$ and low temperature $\beta J = 100$, with $B/J =0.1$ (gapped, left) and $0.2$ (gapless, right). We plot the SC order parameter $\Delta$, charge density $\mathcal{Q}$, magnetization $M$ and entropy density $\mathcal{S}$ as a function of $\mu$. The superconductors are compressible, until they disappear through first-order phase transitions.}
    \label{fig7}
\end{figure}

We finally consider the effect of tuning the chemical potential away from charge neutrality, $\mu \neq 0$ in Eq.~(\ref{eq:H_two_TR_cSYKS}), using the generalized saddle-point equations Eqs.~\ref{eq:SD_equations_tau} and \ref{eq:SD_equations_omega}. In Fig.~\ref{fig7} we show how various physical quantities evolve as a function of $\mu$.  We find that both SC phases are compressible, with a tunable charge density as a function of $\mu$. At larger $\mu$ the system undergoes first-order phase transitions to non-SC phases. While the gapped SC directly transitions to a trivial phase with $\mathcal{Q}=1/2$, the gapless SC first transitions to a non-Fermi liquid with roughly half of the SYK residual entropy, reminiscent of the ``small black hole" phenomenology of Ref.~\cite{Sahoo2020}.

\end{document}